\documentclass[12pt]{article} 
\usepackage{psfig}
\usepackage{a4}
\usepackage{latexsym}

\newcommand{\bra}{\langle}
\newcommand{\ket}{\rangle}
\newcommand{\half}{\frac{1}{2}}

\newcommand{\im}{\mbox{Im}\,}
\newcommand{\re}{\mbox{Re}\,}
\newcommand{\Tr}{\mbox{Tr}\,}
\newcommand{\cl}{{\rm cl}}
\newcommand{\rmd}{{\rm d}}

\newcommand{\lm}{\lambda}

\newcommand{\om}{\omega}  
\newcommand{\be}{\begin{equation}}
\newcommand{\ee}{\end{equation}}
\newcommand{\bea}{\begin{eqnarray}}
\newcommand{\eea}{\end{eqnarray}}
\newcommand{\bean}{\begin{eqnarray*}}
\newcommand{\eean}{\end{eqnarray*}}

\newcommand{\veck}{{\mathbf k}}
\newcommand{\vecp}{{\mathbf p}}
\newcommand{\vecq}{{\mathbf q}}
\newcommand{\vecr}{{\mathbf r}}
\newcommand{\vecx}{{\mathbf x}}

\newcommand{\vecz}{{\mathbf z}}
\newcommand{\vecnul}{{\mathbf 0}}
\newcommand{\ba}{\begin{eqnarray}}
\newcommand{\ea}{\end{eqnarray}}

\newcounter{saveeqn}

\begin{document}
\title{
\vskip -100pt
{  
\begin{normalsize}
\mbox{} \hfill \\
\mbox{} \hfill HD-THEP-01-33\\
\mbox{} \hfill hep-ph/0108125\\
\mbox{} \hfill August 2001\\
\vskip  70pt
\end{normalsize}
}
Spectral function at high temperature in the classical approximation
\author{
Gert Aarts\thanks{email: aarts@thphys.uni-heidelberg.de}
\addtocounter{footnote}{1}\\
\normalsize
{\em Institut f\"ur theoretische Physik, Universit\"at Heidelberg}\\
\normalsize
{\em Philosophenweg 16, 69120 Heidelberg, Germany}\\
\normalsize
}
}
\date{August 15, 2001}
\maketitle
 
\renewcommand{\abstractname}{\normalsize Abstract} 
\begin{abstract}
\normalsize 

At high temperature the infrared modes of a weakly coupled quantum field
theory can be treated nonperturbatively in real time using the classical
field approximation. We use this to introduce a nonperturbative approach
to the calculation of finite-temperature spectral functions, employing the
classical KMS condition in real time. The method is illustrated for the
one-particle spectral function in a scalar field theory in $2+1$
dimensions. The result is compared with resummed
two-loop perturbation theory and both the plasmon mass and width are found
to agree with the analytical prediction.

\vspace{1cm}


\end{abstract}

\newpage

\section{Introduction}

Finite-temperature field theory has received considerable attention during
recent years (see \cite{LeBellac} for a comprehensive textbook). An
important motivation is the physics of the quark-gluon plasma, currently
under investigation at RHIC, as are baryogenesis and reheating after
inflation in the early universe.  Thermal field theory also provides a
necessary reference point for the more complicated case of nonequilibrium
quantum fields.

In thermal equilibrium a prominent role is played by spectral functions
since other correlators can be recovered from it via the
Kubo-Martin-Schwinger (KMS) periodicity condition \cite{KMS}. Thermal
field theory problems can therefore be reduced to a calculation of the
appropriate spectral function. In particular the one-particle spectral
function contains information on the quasiparticle structure of the
theory, needed to describe transport properties of hot matter in the
quark-gluon plasma with the help of a Boltzmann equation
\cite{Blaizot:2001nr}. For the calculation of transport coefficients,
such as the shear viscosity, the presence of a medium-dependent finite
width in the one-particle spectral function is crucial
\cite{Jeon:1993kk}. Resummed perturbative descriptions of the equation of
state of the hot QCD plasma may require a consistent inclusion of
nontrivial quasiparticle spectral functions \cite{Andersen:1999fw}.

In spite of the apparent importance of spectral functions,
nonperturbative computational schemes are rather scarce. A
first-principle approach is offered by lattice field theory. However, the
necessity to use a euclidean formulation hinders access to dynamical
quantities such as spectral functions and other real-time correlators.
Experience in the recovery of mesonic spectral functions in QCD from
euclidean-time correlators has been gained in the last few years using
the Maximum Entropy Method (MEM) \cite{deForcrand:1998}. At high
temperature this approach becomes especially difficult due to the
compactness of the euclidean-time direction. A formulation directly in
real time avoids these problems. Unfortunately a fully nonperturbative
approach to real-time quantum field correlators is still
lacking.\footnote{Recent progress in nonequilibrium quantum field
dynamics using the $2PI$ effective action, including a calculation of the
out-of-equilibrium spectral function for a scalar field in $1+1$
dimensions, can be found in \cite{Berges:2000ur}.}

At high temperature and weak coupling a nonperturbative approach to
real-time quantities is provided by the classical approximation,
originally proposed in the context of high-temperature sphaleron
transitions and electroweak baryogenesis \cite{Grigoriev:1988bd}. Indeed,
at high enough temperature the infrared sector of a thermal field theory
behaves classically as can be guessed from the Bose-Einstein distribution
function at low spatial momenta:
\be 
\label{eq1}
n(\om_\vecp) =
\frac{1}{\exp(\hbar\om_\vecp/T)-1} \to \frac{T}{\hbar\om_\vecp} =
n_\cl(\om_\vecp), \;\;\;\;\;\; \hbar\om_\vecp \ll T, 
\ee 
with $\om_\vecp =\sqrt{\vecp^2+m^2}$ and $T$ the temperature (we take
$\hbar=1$ from now on).  As is well known, a proper definition of
classical thermal field theory requires an inherent ultraviolet cutoff,
provided for instance by a spatial lattice, to regulate the
Rayleigh-Jeans divergence. The importance of the interplay between the
ultraviolet lattice modes and the physical infrared modes has been
realized first in Ref.~\cite{Bodeker:1995pp}. Much progress in the
understanding of the classical approximation and quantum and classical
thermal field theory has been made subsequently, both numerically
\cite{Ambjorn:1995xm} and analytically
\cite{Arnold:1997dy,Aarts:1997qi,Aarts:1998kp}, culminating in
B\"odeker's effective theory for hot infrared nonabelian field dynamics
\cite{Bodeker:1998hm}. A recent review discussing various aspects of the
classical approximation can be found in Ref.~\cite{Bodeker:2001pa}.

In this paper we introduce a nonperturbative approach to the calculation
of spectral functions using the classical field approximation (Sec.~2).
We demonstrate the method with a calculation of the one-particle spectral
function in a scalar field theory in $2+1$ dimensions in Sec.~3.  In
Sec.~4 we calculate the resummed perturbative spectral function and
contrast it with the nonperturbative numerical result. Our findings are
summarized in Sec.~5. For a discussion of the classical analogue of
thermal field theory for a weakly coupled scalar field we refer to  
Ref.~\cite{Aarts:1998kp}.

\section{Classical approximation}

We consider an arbitrary bosonic operator $O$ and define the spectral
function as $i$ times the expectation value of the commutator  
\be
\label{eqspec}
\rho(x-y) = i\bra [O(x),O^\dagger(y)]_-\ket.
\ee 
The brackets denote expectation values at finite temperature $T$, 
\be
\bra O \ket = \frac{1}{Z}\Tr e^{-H/T} O, \;\;\;\;\;\;
Z=\Tr e^{-H/T},
\ee
where the trace is taken over the Hilbert space. Operators $O(x) =
O(t,\vecx)$ are time dependent with the time evolution determined by the
hamiltonian $H$, 
\be
O(t,\vecx) = e^{iHt}O(0,\vecx) e^{-iHt}.
\ee
The spectral function obeys  $\rho^\dagger(x)=-\rho(-x)$ and is, in our
convention, real for an hermitian operator: 
\be 
O^\dagger = O \;\;\;\;\rightarrow \;\;\;\;\rho^\dagger(x)=\rho(x).
\ee
In equilibrium two-point functions depend on the relative coordinates
only and it is convenient to go to momentum space,
\be
\rho(p) = \int \rmd^4x\, e^{-ip\cdot x} \rho(x),
\ee
where $p\cdot x = -p^0x^0+\vecp\cdot\vecx$, $p^0=\om=E$ and $x^0=t$.
We find that in momentum space $\rho(p) \equiv i\rho_{\rm im}(p)$ is
purely imaginary. The imaginary part obeys $p^0\rho_{\rm
im}(p^0,\vecp)>0$ and $\rho_{\rm im}(-p)=-\rho_{\rm im}(p)$. 

For a straightforward discussion of the classical approximation it is
convenient \cite{Aarts:1998kp} to introduce also the Keldysh or
statistical two-point function \cite{Keldysh:1964ud}, 
\be
\label{eqkeldysh}
F(x-y) = \half\bra [O(x),O^\dagger(y)]_+\ket,
\ee 
obeying $F^\dagger(x)=F(-x)$, $F^\dagger(p) = F(p)$. In terms of the usual
Wightman functions \cite{LeBellac}, 
\be
G^>(x-y) = \bra O(x)O^\dagger(y)\ket, \;\;\;\;\;\;
G^<(x-y) = \bra O^\dagger(y)O(x)\ket,
\ee
the spectral and statistical two-point functions read 
\be
\rho(x) = i\left[G^>(x) - G^<(x)\right],\;\;\;\;\;\;
F(x) = \frac{1}{2}[G^>(x) + G^<(x)].
\ee 

In equilibrium the importance of the spectral function is manifest since 
all two-point functions introduced above can be expressed in it,
due to the KMS condition. We find in particular \cite{Aarts:1998kp}
\be
\label{eqKMS}
F(p) = -i\left[ n(p^0)+\half\right]\rho(p),\;\;\;\;\;\;
n(p^0) = \frac{1}{\exp (p^0/T) -1},
\ee 
where $n(p^0)$ is the Bose-Einstein distribution function. This relation
is exact. The question is whether the spectral function can be computed
nonperturbatively.

The classical approximation allows access to nonperturbative 
correlation functions in real time, both in and out of thermal
equilibrium. 
In a classical theory operators commute and the basic classical 
equilibrium correlation function in a bosonic (scalar) theory is given by 
\be
\label{eqS}
S(x-y) = \bra O(x)O^\dagger(y)\ket_\cl \equiv \frac{1}{Z_\cl} \int D\pi
D\phi\,e^{-H/T} O(x)O^\dagger(y),
\ee
with the classical partition function $Z_\cl = \int D\pi
D\phi\,\exp(-H/T)$ and $O(x) = O[\phi(x),\pi(x)]$. This correlator is the
classical equivalent of the Keldysh two-point function
(\ref{eqkeldysh}).  The functional integral is over classical phase-space
at some (arbitrary) initial time,
\be
\int D\pi D\phi = \int\prod_\vecx \rmd\pi(\vecx)\rmd\phi(\vecx),
\ee
weighted with the Boltzmann weight, providing initial conditions
$\phi(0,\vecx)= \phi(\vecx)$ and $\pi(0,\vecx)=\pi(\vecx)$. The
subsequent time evolution is determined from Hamilton's equations of
motion for $\phi(x)$ and $\pi(x)$. The definitions are given for a scalar
field theory, but they can easily be carried over to (non)abelian gauge
theories \cite{Ambjorn:1995xm}. The most convenient formulation employs
the temporal gauge, with the Gauss constraint imposed on the initial
conditions. It is subsequently preserved by the classical equations of
motion.

The classical spectral function is obtained by replacing $-i$ times the
commutator in Eq.~(\ref{eqspec}) with the classical Poisson brackets,
\be
\label{eqspeccl}
\rho_\cl(x-y) = -\bra \{O(x),O^\dagger(y)\}\ket_\cl,
\ee 
defined with respect to the initial fields
\be
\{ A(x), B(y) \} = \int \rmd^3z \left[
 \frac{\delta A(x)}{\delta\phi(\vecz)}
 \frac{\delta B(y)}{\delta\pi(\vecz)}
-\frac{\delta A(x)}{\delta\pi(\vecz)}
 \frac{\delta B(y)}{\delta\phi(\vecz)}\right].
\ee
Due to the formal correspondence between commutators and Poisson brackets 
the quantum and classical spectral function obey the same basic
properties.

At first sight a calculation of the classical
spectral function from the definition in terms of the Poisson bracket
appears rather hard. Fortunately, in thermal equilibrium we may use the
KMS condition to simplify the procedure.
The classical KMS condition is based on the same principle as the usual
KMS condition in a quantum theory: the thermal Boltzmann weight
and the time evolution are controlled by the same hamiltonian
$H$. An easy way to find the classical KMS condition is to consider the
high-temperature (or $\hbar\to 0$) limit of the quantum KMS condition. The
classical equivalent of Eq.~(\ref{eqKMS}) reads (compare Eq.~(\ref{eq1}))
\be
S(p) = -i\frac{T}{p^0}\rho_\cl(p).
\ee
One may also derive this relation directly in the classical
theory without reference to the quantum case
\cite{Parisi,Aarts:1998kp}. This leads to the
classical KMS condition formulated in real space,
\be
\label{eqKMSCL}
\rho_\cl(t,\vecx) = -\frac{1}{T}\partial_t S(t,\vecx).
\ee
This relation will form the basis of the remainder of this paper. It
allows us to calculate the spectral function in real time from a
relatively easily accessible unequal-time correlation function.

\section{Scalar field on the lattice}

As an example we discuss the simple case of a real scalar field $\phi$ in
$2+1$ dimensions with the hamiltonian
\be
H = \int \rmd^2x \left[ \half\pi^2 +  
\half(\nabla\phi)^2 
+ \half m^2\phi^2 + \frac{\lambda}{4!}\phi^4\right].
\ee
We focus on the one-particle spectral function and take
$O(x)=O^\dagger(x)=\phi(x)$.
Using Eqs.~(\ref{eqKMSCL}) and (\ref{eqS}) we find directly
\be
\label{eq18}
\rho_\cl(t,\vecx) = -\frac{1}{T}\bra
\pi(t,\vecx)\phi(0,\vecnul)\ket_\cl.
\ee
The right-hand-side of Eq.~(\ref{eq18}) can be computed numerically in a
straightforward manner, as follows. The classical field theory is
defined on a spatial lattice with $N\times N$ sites and lattice
spacing $a$ and we use periodic boundary conditions. To solve the
dynamics we use a leapfrog discretization with time step $a_0<a$. The
canonical momenta $\pi(t+\half a_0,\vecx)$ are defined at intermediate
time steps, which suggests to use a symmetrized definition of the
spectral function on the lattice
\be
\label{eq19}
\rho_{\cl,\rm lat}(t,\vecx) = -\frac{1}{T}\left\bra
\pi(t+\half a_0,\vecx)\half\left[\phi(0,\vecnul)+\phi(a_0,\vecnul)\right] 
\right\ket_\cl.
\ee
\begin{figure}
\centerline{\psfig{figure=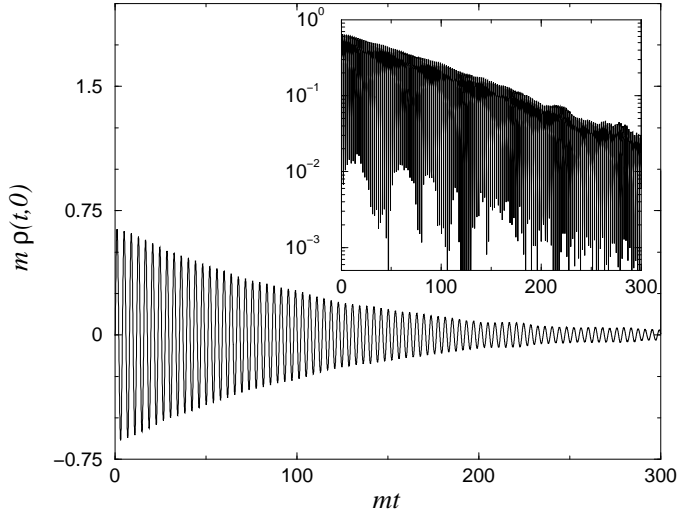,height=7cm}}
\caption{Classical spectral function $\rho_\cl(t,\vecnul)$ at zero
momentum in real time. The inset shows the absolute value on a log scale.
The temperature is $T/m=7.2$. Without loss of generality we use $\lm/m=1$ 
throughout. 
}
\label{fig1}
\end{figure} 
To generate thermal initial conditions we use the Kramers equation 
algorithm, a variant of the hybrid Monte Carlo method 
\cite{Jansen:1995gz}. The evolution in real time is calculated using
classical equations of motion. In a simulation we switch therefore
between noisy evolution to create independent thermal configurations and
hamiltonian evolution to calculate observables. The results presented
below are obtained using 2000 independently thermalized initial
configurations for each temperature. The mass scale $m$ is used as the
dimensionful scale and the results presented are obtained with
$N=128$, $am=0.2$ and $a_0/a=0.1$ (note that the finite time
step affects the equal-time canonical
relation $\partial_t\rho_\cl(t,\vecx)|_{t=0}=\delta(\vecx)$).
In a classical theory the coupling constant $\lambda$ can be scaled out of
the equations of motion and the remaining dimensionless combination is
$\lambda T/m^2$ (recall that $\lambda$ has a dimension of mass in
$2+1$ dimensions). Without loss of generality we take therefore 
$\lambda/m=1$ throughout. Larger $T$ corresponds then to a larger
effective interaction strength. In the simulations the temperature is
determined from the average kinetic energy $T = 
a^2\bra\pi^2(t,\vecx)\ket_\cl$ and the temperatures we consider are such
that $aT = {\cal O}(1)$.

\begin{figure}
\centerline{\psfig{figure=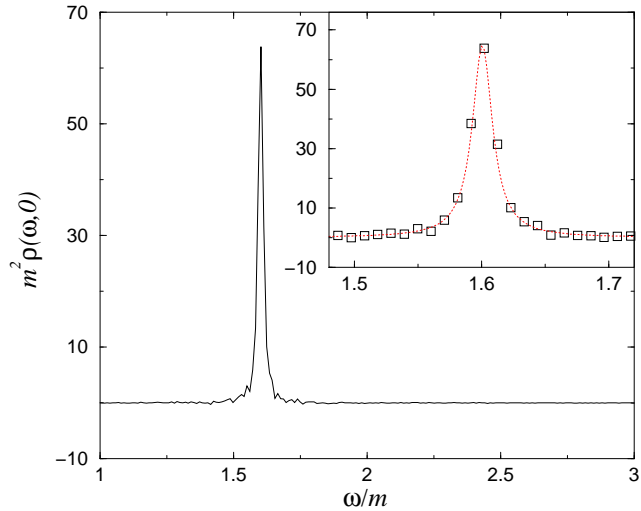,height=7cm}}
\caption{Spectral function $\rho_\cl(\om,\vecnul)$, obtained from
the real-time result of Fig.~\ref{fig1} by a sine-transform, versus
the frequency $\om$. The inset shows a magnification with the data
points indicated. The dotted line is a fit to a Breit-Wigner function.  
}
\label{fig2}
\end{figure} 

In Fig.~1 we present the classical spectral function at zero spatial
momentum, obtained from a volume average of Eq.~(\ref{eq19}), in real
time at a temperature $T/m=7.2$. We see oscillating approximately
exponentially damped behaviour.  The spectral function in frequency space
can be obtained from a sine-transform 
\be
\label{eq20}
\rho_{\cl}(\om,\vecp) = 2\int_0^{t_{\rm max}} \rmd t\,
\rho_{\cl, \rm lat}(t,\vecp)\sin\om t,
\ee
where we used the antisymmetry of the spectral function under time
reflection and absorbed the $i$ of the previous section directly in the
definition. The result is shown in Fig.~2. The spectral
function consists of a single peak with the narrow width.
The inset shows that the peak is well described by a 
Breit-Wigner spectral function (see Sec.~4). We have tried to find other
contributions at larger frequency, but these could not be seen in the
numerical data.
In the numerical simulation the late-time regime becomes
more and more difficult to establish, resulting typically in wiggly
nondamped behaviour. This limits the time interval that can be used in
the sine-transform to a maximal time $t_{\rm max}$ and constrains the
resolution in frequency-space to $\Delta\om = \pi/t_{\rm max}$. For the
result in Fig.~2 the resolution is $\Delta\om/m=\pi/300\approx 0.01$, as
can be seen from the inset.

\section{Perturbative expectation}

The spectral function in the quantum theory can be expressed in terms
of the retarded self energy $\Sigma_R =\re \Sigma_R +i\,\im \Sigma_R$ as
\cite{Weldon:1983jn,Wang:1996qf}
\be
\rho_{\rm im}(\om,\vecp) = \frac{-2\im
\Sigma_R(\om,\vecp)}{
\left[\om^2-\vecp^2-m^2-\re\Sigma_R(\om,\vecp)\right]^2 +
\left[\im \Sigma_R(\om,\vecp)\right]^2}.
\ee
The theory exhibits a quasiparticle structure, with the quasiparticle
often referred to as the plasmon, in the limit that the rate
$\Gamma(\om,\vecp) = -\im \Sigma_R(\om,\vecp)/\om$
is much smaller than $(\vecp^2+m^2+\re\Sigma_R)^{1/2}$. In this case the
spectral function is well approximated with a Breit-Wigner function,
which at zero momentum reads 
\be
\rho_{\rm BW}(\om,\vecnul) =
\frac{2\om\Gamma}{(\om^2-M^2)^2+\om^2\Gamma^2}.
\ee
Here $M$ is the plasmon mass and $\Gamma$ its width (at zero momentum)
\be
\Gamma = -\frac{\mbox{Im}\, \Sigma_R(M,\vecnul)}{M}.
\ee
In this limit contributions from multiparticle states beyond the
three-particle threshold are tiny. 

A perturbative calculation of the retarded self energy is
standard in thermal field theory (see e.g.\ \cite{Wang:1996qf} for a clear
discussion in $3+1$ dimensions). For the $(2+1)$-dimensional
case we consider here we find the following.
At one-loop order the tadpole diagram 
\be
\Sigma^{(1)}_R = \frac{\lambda}{2} \int
\frac{\rmd^2p}{(2\pi)^2}\frac{n(\om_\vecp)+\half}{\om_\vecp}
\ee
contributes to the mass shift only. Resummation of the tadpole diagram in
the limit of high temperature and weak coupling results in a gap
equation for the resummed mass parameter $M$:
\be
\label{eqM}
M^2 = \frac{\lambda T}{4\pi} \log \frac{T}{M}\;\;\;\;\;\;\mbox{ (one-loop
resummed)},
\ee
where the zero-temperature mass is neglected.
In $2+1$ dimensions the one-loop mass is sensitive 
to both the ultraviolet momentum scale (cutoff by $T$) and the infrared
momentum scale (cutoff by $M$).
A finite width in the spectral function arises at two-loop
order from the imaginary part of the setting-sun diagram. 
We focus here on on-shell $2\to 2$ scattering for which the contribution
reads 
\bean
\mbox{Im}\, \Sigma^{(2)}_R(\om_\vecp,\vecp) &=& 
-\frac{\lambda^2}{4}\int \rmd\Phi_{123}(\vecp)\,
2\pi\delta(\om_\vecp +\om_\veck-\om_\vecq-\om_\vecr) \\
&&\times  \left\{
n(\om_\veck)[1+n(\om_\vecq)][1+n(\om_\vecr)] - 
[1+n(\om_\veck)]n(\om_\vecq)n(\om_\vecr)\right\},  
\eean  
with 
\begin{figure}
\centerline{\psfig{figure=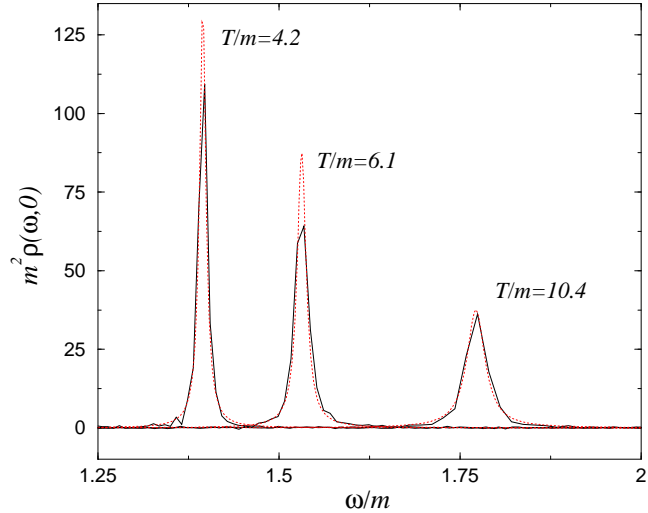,height=7cm}}
\caption{
Spectral functions $\rho_\cl(\om,\vecnul)$ for various temperatures
$T$. Fits to a Breit-Wigner function are shown with dotted lines. The
height, position and width of the peak are approximately related as
$\rho_\cl(M,\vecnul) \sim 2/M\Gamma$.
}
\label{fig3}
\end{figure} 
\be
\rmd\Phi_{123}(\vecp) =
\frac{\rmd^2k}{(2\pi)^22\om_\veck}
\frac{\rmd^2q}{(2\pi)^22\om_\vecq}
\frac{\rmd^2r}{(2\pi)^22\om_\vecr}
(2\pi)^2\delta(\vecp-\veck-\vecq-\vecr).
\ee  
The on-shell dispersion relations contain the one-loop resummed mass
parameter, $\om_\vecp=\sqrt{\vecp^2+M^2}$. It is easy to check that the
momentum integrals are dominated by the infrared modes. These soft modes
are classical. The leading contribution at high temperature
and weak coupling can therefore be obtained by replacing $n(\om) \to T/\om
=n_\cl(\om)$. At zero momentum $\vecp=0$ the integrals can be performed
analytically and after a straightforward calculation we
find the plasmon width in $2+1$ dimensions to be
\be 
\label{eqGamma}
\Gamma = c\frac{\lambda^2T^2}{M^3},
\;\;\;\;\;\;c=\frac{3-2\sqrt{2}}{32\pi}
\approx 0.00171.
\ee

In Fig.~3 the classical spectral function and fits to a Breit-Wigner
function are shown for various temperatures. From the fits one may
extract estimates for the classical plasmon mass and width. This provides
a possibility to compare nonperturbatively determined classical plasmon
masses and widths with perturbative calculations and address the
applicability of one-loop resummed perturbation theory at finite
temperature.

\begin{figure}[t]
\centerline{\psfig{figure=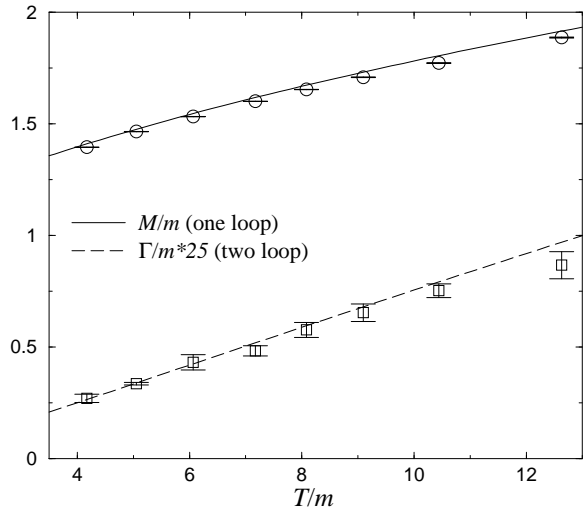,height=7cm}}
\caption{Mass $M_\cl$ (circles) and width $\Gamma_\cl$ (squares) of the
zero-momentum spectral function as a function of the temperature. The
width is multiplied with 25 for clarity. The data points are
obtained from fits to a Breit-Wigner function, with the statistical error
estimated from a jackknife analysis. The lines are the predictions for
$M_\cl$ from the one-loop gap equation (full) and $\Gamma_\cl$ from the
on-shell two-loop contribution (dashed).
}
\label{fig4}
\end{figure} 

The calculations in the quantum theory can easily be carried over to the
classical approximation used in the numerical calculation
\cite{Aarts:1998kp}.  The one-loop gap equation for the classical mass
parameter $M_\cl$ reads
\be
M_\cl^2 = m^2 +\frac{\lambda}{2} \int
\frac{\rmd^2p}{(2\pi)^2}\frac{T}{\vecp^2+M_\cl^2}.
\ee
The integral is logarithmically divergent in the ultraviolet, which 
reflects the $\log T$ contribution in Eq.\ (\ref{eqM}) in the quantum
theory. In the classical theory the divergence is regulated by the
lattice cutoff. If desired, it can be matched (renormalized) by adjusting
the mass parameter $m$ in the effective classical theory
\cite{Aarts:1997qi}. 
In order to compare the one-loop and the nonperturbative calculation we
write the gap equation on the lattice, 
\be
M_\cl^2 = m^2 +\frac{\lambda T}{2}\frac{1}{L^2} \sum_{n_1,n_2}
\frac{1}{\hat\vecp^2  +M_\cl^2}, \;\;\;\;\;\;\hat\vecp^2 = \sum_{i=1}^2
\frac{2}{a^2}(1-\cos ap_i),
\ee
with $p_i=2\pi n_i/L_i$, $-N_i/2+1 \leq n_i \leq N_i/2$ and $L_i=aN_i=L$
($i=1,2$), and solve it numerically.\footnote{
In the infinite volume limit the lattice gap equation can be written as
$M_\cl^2 = m^2 + \lambda T/(4\pi)k F(\pi/2, k)$, with $1/k =
1+a^2M_\cl^2/4$ and $F(\pi/2,k)$ the complete elliptic function of the
first kind.
}
The result is presented in Fig.~4 for various temperatures. Maybe
surprisingly we see that the nonperturbative determination of the plasmon
mass and the resummed one-loop result differ at most a few percent,
indicating the relative unimportance of higher-loop contributions
for this quantity.  Note that for our parameters the effective mass
$M_\cl$ is small in lattice units ($aM_\cl \sim 0.3$).

An analytical expression for the width of the spectral function in the
classical approximation can be obtained directly from the calculation
carried out above.  As indicated, the perturbative width of the spectral
function in the quantum theory is dominated by soft momenta and coincides
therefore with the classical result to leading order: $\Gamma_\cl$ is
given by Eq.\ (\ref{eqGamma}) after the replacement $M\to M_\cl$.  A
comparison between the perturbatively and nonperturbatively determined
widths is presented in Fig.~4 as well. Again agreement between
the two calculations can be seen, indicating that the dominant
contribution is produced by the lowest order two-loop result.
We emphasize that the width is a typical real-time quantity and
therefore not easily accessible by other nonperturbative methods.

\section{Summary}

We have used the classical field approximation at high temperature and
weak coupling to formulate a nonperturbative method for the calculation
of spectral functions at finite temperature, with the help of the
classical KMS condition. We focused on the one-particle spectral function
in a scalar field theory in $2+1$ dimensions. For the temperatures
investigated our numerical results indicate that the one-particle
spectral function is a simple narrow peak: the spectral function is
completely dominated by the plasmon.  To compare with perturbation theory
we calculated the one-loop resummed plasmon mass and the two-loop
contribution to the width. Agreement between the nonperturbative
numerical simulation and the perturbative expressions was found. These
results provide a justification for the use of resummed perturbation
theory for a weakly coupled scalar field in equilibrium as well as for
kinetic approaches based on two-loop approximations close to equilibrium.
	
It would be interesting to extend the analysis to more complicated
spectral functions. Transport coefficients can be expressed in terms of
spectral functions of composite operators. These are in general more
sensitive to the ultraviolet scale and therefore to the Rayleigh-Jeans
divergence in a classical limit. Nevertheless, it might be interesting to
test perturbative ideas against a nonperturbative numerical calculation.
When sufficient care concerning the gauge symmetry is taken, methods
applied here might also offer the possibility to gain further insight in
nonperturbative aspects of hot gauge plasmas.

\vspace{5mm}

\noindent
{\bf Acknowledgements}

I would like to thank I.-O.~Stamatescu for discussions and J.~Berges for
collaboration on related topics. 
This work was supported by the TMR network {\em Finite
Temperature Phase Transitions in Particle Physics}, EU contract no.\
FMRX-CT97-0122.

\end{document}